\documentclass{article}
\usepackage{arxiv}
\usepackage[utf8]{inputenc} 
\usepackage{hyperref}       
\usepackage{amsfonts}       
\usepackage{amsmath}
\usepackage{color}
\usepackage{graphicx}
\usepackage{subfigure}
 \usepackage{cite}
\usepackage[title]{appendix} 

\title{An algorithm for exact analytical solutions for tilted anisotropic Dirac materials}
\author{Julio A. Mojica-Zárate$^{*,1}$, Daniel O-Campa$^{\dagger,2}$ and Erik Díaz-Bautista$^{\dagger,3}$\\
$^{\dagger}$Unidad Profesional Interdisciplinaria de Ingeniería Campus Hidalgo \\
del Instituto Politécnico Nacional, 42162 Hidalgo, México.\\
$^{*}$Área Académica de Matemáticas y Física\\
Universidad Autónoma del Estado de Hidalgo, 42184 Hidalgo, México\\
e-mail: mo318293@uaeh.edu.mx$^1$, dortizca@ipn.mx$^2$, ediazba@ipn.mx$^3$}
\begin{document}
\maketitle
\begin{abstract}
In this article, we obtain the exact solutions for bound states of tilted anisotropic Dirac materials under the action of external electric and magnetic fields with translational symmetry. In order to solve the eigenvalue equation that arises from 
the effective Hamiltonian of these materials, we describe an algorithm that allow us to decouple 
the differential equations that are obtained for the spinor components.
\end{abstract}
\section{Introduction}
Graphene was experimentally isolated by Geim and Novoselov in 2004 \cite{novoselov2004} and since then, due to its electric properties such as the integer quantum Hall effect, plenty of experimental and theoretical research has been focused on to discover materials with similar or improved electronic properties \cite{Feng2016,Feng20161,Zhao2018,Tajima2009,Goerbig2008,Goerbig2009,Morinari2009,Sabsovich2020,Schaibley2016,Ang2017}. Some of these materials are borophene, Weyl semimetals, dichalcogenides, the organic conductor $\alpha$-(BEDT-TTF)$_2$I$_3$ among others. It is known that the charge carriers in such materials behave as massless chiral quasiparticles with a linear dispersion relation at low energies, leading to a description in terms of a Dirac-like effective Hamiltonian.

In general, the lattice structure of these materials is not as simple as that of graphene. As a result, tilted anisotropic Dirac cones are presented in their band structure so that the electronic and transport properties are valley dependent. As happens with graphene, these materials could be used in the design of electronic devices. For that reason, the study of the interaction between their electrons and external fields becomes relevant in order to find a way to confine the charged particles taking into account the valley dependency. 

In this article, we will consider the interaction of electrons in anisotropic Dirac materials with external magnetic and electric fields whose strength varies along the $x$ axis on the plane of the sample ($x-y$). This physical configuration is similar to the one addressed for monolayer graphene and allows us to make use of the translation invariance to find the Landau levels and their corresponding eigenfunctions\cite{Kuru2009,concha2018}.

This paper is organized as follows: in section \ref{sec2} we will introduce the effective Hamiltonian that describes tilted anisotropic Dirac materials under the action of external electric and magnetic fields and present the corresponding eigenvalue problem when a translational symmetry is assumed. Section \ref{sec3} contains the algorithm that allows us to find the eigenstates in terms of the eigenvectors and eigenvalues of a $2\times2$ matrix. In section \ref{sec4} we study some examples of magnetic and electric fields to provide an exactly solvable problem through the method developed and we present a discussion of the results. Our conclusions are contained in section \ref{sec5}
\section{The model}\label{sec2}
At low energies, the Dirac materials that exhibit tilted cones are described by the effective Hamiltonian in natural units ($\hbar=e=c=1$)
\begin{equation}
\mathcal{H}_0=\nu\left(v_x p_x \sigma_x + v_y p_y\sigma_y+ v_t p_y \sigma_0\right),
\label{1}
\end{equation}
being $\nu=\pm 1$ the valley index (when $\nu=1$, we will refer to valley K, while $\nu=-1$, to valley K'), $v_{x}$ and $v_{y}$ are the anisotropic Fermi velocities and $v_{t}$ is a term of velocity that arises from the tilting of the Dirac cones, $\sigma_{x,y}$ denote the Pauli matrices and $\sigma_{0}$ is the $2\times2$ identity matrix. In general, velocities $v_x$, $v_y$ and $v_t$ depend on the material under study.

Considering now the interaction with a magnetic field $\vec{\mathcal{B}}$ perpendicular to the surface of the material ($x-y$ plane) and an in-plane electric field $\vec{\mathcal{E}}$, then the Hamiltonian in Eq. \eqref{1} is modified by the minimal coupling rule as
\begin{equation}
\mathcal{H}=\nu\left[v_x (p_x+\mathcal{A}_x(x,y)) \sigma_x + v_y (p_y+\mathcal{A}_y(x,y))\sigma_y+ v_t (p_y+\mathcal{A}_y(x,y)) \sigma_0\right]-\phi(x,y)\sigma_0,
\label{2}
\end{equation}
where $\mathcal{A}_i(x,y)$ ($i=x,y$) represents the components of the vector potential $ \vec{\mathcal{A}}(x,y)$ and $\phi(x,y)$ denotes the scalar potential, such that $\vec{\mathcal{B}}=\nabla \times \vec{\mathcal{A}}(x,y)$ and $\vec{\mathcal{E}}=-\nabla \phi(x,y)$. It is important to note that for both Hamiltonians in Eqs. \eqref{1} and (\ref{2}), the probability and current densities are given by
\begin{equation}
\rho(x,y,t)=\Psi^{\dagger}\Psi,\quad\vec{\mathcal{J}}(x,y,t)=\Psi^{\dagger}\vec{j}\Psi,  
\label{3}
\end{equation}
being $\Psi=\Psi(x,y,t)$ an arbitrary state while the components of $\vec{j}$ turning out to be
\begin{equation}
j_x=\nu v_x \sigma_x, \quad 
j_y=\nu \left( v_y \sigma_y+v_t \sigma_0\right).
\label{4}
\end{equation}
We have to remark that both potentials in Eq. (\ref{2}) only depend on the coordinates in the plane. Next, we will discuss the eigenvalue problem for the above Hamiltonian giving some considerations about applied external fields.
\subsection{The eigenvalue problem}
Let us start by defining the following quantities in order to solve the eigenvalue problem:
\begin{equation}
x_c=\sqrt{\frac{v_y}{v_x}} x, \quad 
p_x^c=\sqrt{\frac{v_x}{v_y}} p_x, \quad 
y_c=\sqrt{\frac{v_x}{v_y}} y, \quad
p_y^c=\sqrt{\frac{v_y}{v_x}} p_y,
\nonumber
\end{equation}
\begin{equation}
\mathcal{A}_x^c(x_c,y_c)=\sqrt{\frac{v_x}{v_y}}\mathcal{A}_x(x,y),\quad\mathcal{A}_y^c(x_c,y_c)=\sqrt{\frac{v_y}{v_x}}\mathcal{A}_y(x,y),\quad\quad\phi_c(x_c,y_c)=\sqrt{\frac{v_y}{v_x}}\phi(x,y).
\label{5}
\end{equation}
Using the above expressions, the effective Hamiltonian in Eq. \eqref{2} reads as
\begin{equation}
\mathcal{H}= \nu v_{\rm F}\left[\vec{\sigma}\cdot \left(\vec{p_c}+\vec{\mathcal{A}_c}(x_c,y_c)\right)+\frac{v_t}{v_y}\left(p_y^c+\mathcal{A}_y^c(x_c,y_c)-\frac{\nu}{v_t}\phi_c(x_c,y_c)\right)\sigma_0\right],
\label{6}
\end{equation}
where $v_{\rm F}$ represents the geometric mean of $v_x, v_y$\cite{betancur2021,diaz2022time}. The changes proposed in Eq. \eqref{5} preserve the canonical commutation relation between position and momentum operators. In addition, if the material does not present tilted cones and there is no electric field, the quantities in Eq. \eqref{5} give us the advantage of working in new coordinates for which the material can be described, in an effective way, as an isotropic one as it happens with graphene\cite{Kuru2009}. On the other hand, if the strength of the fields varies only in a fixed direction, namely $x$, then the vector potential can be represented in the Landau gauge as $\vec{\mathcal{A}}=\mathcal{A}_y(x)\hat{e}_y$, such that $\vec{\mathcal{B}}=\mathcal{A}_y'(x)\hat{e}_z$, while the scalar one $\phi(x,y)=\phi(x)$ implies that $\vec{\mathcal{E}}=-\phi'(x)\hat{e}_x$. Thus, for stationary states, the time-independent eigenvalue equation reads as
\begin{equation}
\left(\mathcal{H}-E\sigma_0\right)\Psi\left(x,y\right)=0.
\label{7}
\end{equation}
Due to the assumptions about the electric and magnetic fields made previously, $\left[\mathcal{H},p_y\right]=\left[\mathcal{H},p^c_y\right]=0$ holds. Hence, the eigenfunctions can be written as
\begin{equation}
\Psi\left(x,y\right)=e^{ik_c y_c}\bar{\Psi}\left(x_c\right),
\label{8}
\end{equation}
where $\bar{\Psi}\left(x_c\right)=\left(\psi^+\left(x_c\right),i\psi^-\left(x_c\right)\right)^{\rm T}$. Note that the product $k_c y_c$ has to be equal to $k_y y$, where $k_y$ is the wavenumber in $y$ direction and therefore $k_c=(\sqrt{v_y/v_x})k_y$. By substituting Eq. \eqref{8} into Eq. \eqref{7} and taking into account that $p_y^c=-i{\rm d}/{\rm d}x_c$, the eigenvalue equation lead us to the following matrix system
\begin{equation}
\left[-i\frac{\rm d}{{\rm d}x_c}\sigma_x+\left(k_c+\mathcal{A}_y^c(x_c)\right)\sigma_y+\left(\frac{v_t}{v_y}\mathcal{A}_y^c(x_c)-\frac{\nu}{v_y}\phi_c(x_c)-\bar{E}\right)\sigma_0\right]\bar{\Psi}\left(x_c\right)=0,
\label{9}
\end{equation}
where $\bar{E}=(\nu E- v_t k_y)/v_{\rm F}$. By considering an invertible change of variable $z=z(x_c)$, such that $\mathcal{A}_y^c(x_c)=\bar{\mathcal{A}}_y^c(z)$ and $\phi_c(x_c)=\bar{\phi}_c(z)$ for some functions $\bar{\phi}_c$ and $\bar{\mathcal{A}}_y^c$, and multiplying by $i\sigma_x$ to the left-hand side of Eq. \eqref{9}, we get
\begin{equation}
g(z)\frac{{\rm d}\bar{\Theta}\left(z\right)}{{\rm d}z}=\left[i\mbox{F}_1(z)\sigma_x+\mbox{F}_2(z)\sigma_z\right]\bar{\Theta}\left(z\right),
\label{10}
\end{equation}
being $\bar{\Theta}\left(z\right)$ a new two-component spinor that complies with $\bar{\Theta}\left(z\right)=\bar{\Psi}\left(x_c\right)$ and
\begin{equation}
g(z)=\frac{{\rm d}z}{{\rm d}x_c},\quad
\mbox{F}_1(z)=\left(\bar{E}+\frac{\nu}{v_y}\bar{\phi}_c(z)-\frac{v_t}{v_y}\bar{\mathcal{A}}_y^c(z)\right),\quad
\mbox{F}_2(z)=(k_c+\bar{\mathcal{A}}_y^c(z)).
\label{11}
\end{equation}
Notice that, due to the explicit appearance of $\sigma_x$ in Eq. \eqref{10}, the components of the spinor $\bar{\Theta}\left(z\right)$ satisfy a coupled system of differential equations. On the other hand, although the equality $\bar{\Psi}\left(x_c\right)=\bar{\Theta}\left(z\right)$ fulfills, the spinors are not functionally the same. This fact must be taken into account when the normalization process is performed to ensure that
\begin{equation}
\int^{\infty}_{-\infty}
|\Psi\left(x,y\right)|^2{\rm d} x=1.
\label{12}
\end{equation}
As a consequence, the other spinors comply with the following relationship
\begin{equation}
\sqrt{\frac{v_x}{v_y}}\int^{\infty}_{-\infty}
|\bar{\Psi}\left(x_c\right)|^2{\rm d} x_c=
\sqrt{\frac{v_x}{v_y}}\int^{z^+}_{z^-}\frac{|\bar{\Theta}\left(z\right)|^2}{g(z)}
{\rm d} z=1,
\label{13}
\end{equation}
with $z^{\pm}=z(\pm\infty)$. Next, we will describe an algorithm that allows us to decouple the system given in Eq. \eqref{10}, and that guarantees square-integrability of the solution, obtaining in that way the eigenvalues and eigenfunctions of the Hamiltonian $\mathcal{H}$.
\section{The matrix $\mathbb{K}$}\label{sec3}
In order to find the eigenfunctions of the Hamiltonian \eqref{6}, we propose the following procedure that starts by applying the operator $g(z)\,{\rm d}/{\rm d}z$ on both sides of Eq. \eqref{10}. After some algebra, we get 
\begin{equation}
\left(\mathcal{D}_{z}+\mathbb{K}\right)\bar{\Theta}\left(z\right)=0,
\label{14}
\end{equation}
where $\mathcal{D}_{z}$ is a differential operator and $\mathbb{K}$ is a $2\times 2$ matrix given by
\begin{equation}
\mathcal{D}_{z}=g^2(z)\frac{{\rm d}^2}{{\rm d}z^2}+\mbox{F}^2_1(z)-\mbox{F}^2_2(z),\quad
\mathbb{K}=
\begin{pmatrix}
-a_2(z)&-ia_1(z)\\
&\\
-ia_1(z)&a_2(z)
\end{pmatrix},
\label{15}
\end{equation}
being $a_j(z)\equiv \mbox{F}'_j(z)g(z)-\mbox{F}_j(z)g'(z)$ with $j=1,2$. Here $f'(z)$ denotes the derivative with respect to $z$. Let us note that the new system of differential Eqs. in \eqref{14} maintains the coupling between the entries of spinor $\bar{\Theta}(z)$. Nevertheless, if the matrix $\mathbb{K}$ is considered constant, then we can use an auxiliary spinor $\bar{\Phi}(z)$ in the following way:
\begin{equation}
\bar{\Theta}\left(z\right)=
\mathbb{S}
\begin{pmatrix}
\phi^+(z)\\ 
\\
i \phi^-(z)
\end{pmatrix}=\mathbb{S}\bar{\Phi}\left(z\right),
\label{16}
\end{equation}
where $\mathbb{S}=(\vec{v_1}, \vec{v_2})$ is a $2\times2$ matrix whose column vectors $\vec{v_j}$ correspond to the eigenvectors of $\mathbb{K}$, i.e., $\mathbb{K}\vec{v_j}=\lambda_j \vec{v_j}$ for $j=1,2$, and $\lambda_j$ being the corresponding eigenvalues. By plugging the Eq. \eqref{16} into \eqref{14} and multiplying by $\mathbb{S}^{-1}$ on the left-hand side, it turns out to be 
\begin{equation}
\left(\mathcal{D}_{z}\mathbb{S}^{-1}\mathbb{S}+\mathbb{S}^{-1}\mathbb{K}\mathbb{S}\right)\bar{\Phi}\left(z\right)=0.
\label{17}
\end{equation}
Since the columns of matrix $\mathbb{S}$ are constructed from the eigenvectors of matrix $\mathbb{K}$, the similarity transformation $\mathbb{S}^{-1}\mathbb{K}\mathbb{S}$ leads to
\begin{equation}
\left(\mathcal{D}_{z}+\mathbb{M}\right)\bar{\Phi}\left(z\right)=0,
\label{18}
\end{equation}
where $\mathbb{M}\equiv\mathbb{S}^{-1}\mathbb{K}\mathbb{S}$ is the diagonal matrix given by 
\begin{equation}
\mathbb{M}=
\begin{pmatrix}
\lambda_1&0\\
&\\
0&\lambda_2
\end{pmatrix}.
\label{19}
\end{equation}
In this way, the Eq. \eqref{14} is transformed into two eigenvalue equations where the operator $\mathcal{D}_{z}$ acts on the auxiliary functions $\phi^{\pm}(z)$. Explicitly:
\begin{align}
\mathcal{D}_{z}\phi^{+}(z)&=-\lambda_1\phi^{+}(z),\nonumber\\
\mathcal{D}_{z}\phi^{-}(z)&=-\lambda_2\phi^{-}(z),
\label{20}
\end{align}
Here, we must remark on the following points:
\begin{itemize}
\item[•] To satisfy the normalization condition, it is necessary to require that the auxiliary functions $\phi^{\pm}(z)$ be square-integrable, and therefore, they must vanish at the points $z^{\pm}$ specified in Eq. \eqref{13}. Thus, through the auxiliary spinor $\bar{\Phi}(z)$, we can recover $\bar{\Psi}(x_c)$ in \eqref{8}.
\item[•] The entries $a_j$ take real values. Then, the matrix $\mathbb{K}$ is neither hermitian nor normal. Thus, we can not guarantee that $\mathbb{S}$ to be unitary, so the spinor $\bar{\Theta}\left(z\right)$ in Eq. \eqref{16} needs to be properly normalized, even if $\bar{\Phi}(z)$ is.
\item[•] The matrix $\mathbb{S}$ is not unique. Therefore, the matrix $\mathbb{M}$ is not unique either but it can only be $\mathbb{M}=\mbox{diag}(\lambda_1,\lambda_2)$ or $\mathbb{M}=\mbox{diag}(\lambda_2,\lambda_1)$ (see appendix \ref{apendice1}). 
\end{itemize}
On the other hand, if we focus on the entries $a_j$, we can express them as
\begin{equation}
a_j(z)=g^2(z)\frac{{\rm d}}{{\rm d}z}\left(\frac{\mbox{F}_j(z)}{g(z)}\right),\quad
\mbox{or}\quad
a_j(z)=-\mbox{F}_j^2(z)\frac{{\rm d}}{{\rm d}z}\left(\frac{g(z)}{\mbox{F}_j(z)}\right).
\label{21}
\end{equation}
However, by assuming that $\mathbb{K}$ is constant, we can obtain integral expressions for $\mbox{F}_j(z)$ and $g(z)$, namely:
\begin{subequations}
\begin{alignat}{2}
\mbox{F}_j(z)&=a_jg(z)\int\frac{{\rm d}z}{g^2(z)}+c_jg(z),\label{22a}\\
g(z)&=-a_j\mbox{F}_j(z)\int\frac{{\rm d}z}{\mbox{F}_j^2(z)}+\tilde{c}_j\mbox{F}_j(z).
\label{22b}
\end{alignat}
\label{22}
\end{subequations}
where $c_j$ and $\tilde{c}_j$ are constants of integration. It is crucial to emphasize that, although both equations ensure the constancy of the matrix $\mathbb{K}$ and vice versa, each one approaches differently the derivation of the decoupled system of equations presented in \eqref{20}. On one hand, Eq. \eqref{22a} allows us to propose invertible changes of variable, $z(x_c)$, all through the function $g(z)$, and from those, to determine the necessary information about the profiles of the magnetic and electric fields. On the other hand, Eq. \eqref{22b} is based on knowledge of the applied electric and magnetic fields (through the functions $\mbox{F}_j$). Also, it helps us to determine the function $g(z)$ and, consequently, the necessary change of variable. However, it does not guarantee the invertibility of the change of variable $z(x_c)$. Furthermore, we cannot assure that the required relationship between these variables has an analytical form.\\
\\
Therefore, taking into account the disadvantages involved in using Eq. \eqref{22b}, we will analyze the algorithm of the matrix $\mathbb{K}$ described previously, using the function $g(z)$ as a starting point. In this way, we will obtain profiles of electric and magnetic fields that lead to solvable cases, as well as the analytical solutions corresponding to the eigenvalue problem posed in \eqref{7}.
\section{An application of the algorithm}\label{sec4}
Let us start by noting that $z=z(x_c)$ can not be a constant since it represents a change of variable. Thus $g(z)={\rm d}z/{\rm d}x_c$ can not be the zero constant function. However, it is possible to use the matrix $\mathbb{K}$ algorithm to study some cases in which the function $g$ takes a non-null value as we will do below.
\subsection{$g(z)$ as a linear function}
Let us begin by considering that $g$ is defined as $g(z) = d_1 z + d_2$, where $d_1$ and $d_2$ are arbitrary constants, both non-null simultaneously. Assuming that at least $d_1$ is non-zero, we can infer from Eq. \eqref{11} that $z(x_c)$ becomes:
\begin{equation}
z=\mbox{exp}\left[d_1 (x_c+d_3)\right]-\frac{d_2}{d_1},
\label{23}
\end{equation}
being $d_3$ a new constant of integration. On the other hand, using the provided profile of $g$ in Eq. \eqref{22a}, it turns out that
\begin{equation}
\mbox{F}_j(z)=-\frac{a_j}{d_1}
+c_jd_1z+d_2c_j.
\label{24}
\end{equation}
By substituting $\mbox{F}_j$ into Eq. \eqref{11} and taking into account Eq. \eqref{5}, after some algebraic manipulations, we obtain the following expressions for the vector and scalar potentials:
\begin{subequations}
\begin{alignat}{2}
\mathcal{A}_y(x)&=\sqrt{\frac{v_x}{v_y}}\left[c_2d_1\, \mbox{exp}\left(d_1(d_3+\sqrt{\frac{v_y}{v_x}}x)\right)-\frac{a_2+d_1k_c}{d_1}\right],
\label{25a}\\
\phi(x)&=\sqrt{\frac{v_x}{v_y}}\frac{d_1(c_1 v_y+c_2 v_t)\, \mbox{exp}\left[d_1 \left(d_3+x \sqrt{\frac{v_y}{v_x}}\right)\right]-\frac{a_1 v_y+a_2 v_t+d_1 \bar{E} v_y+d_1 k_c v_t}{d_1}}{\nu}.
\label{25b}
\end{alignat}
\label{25}
\end{subequations}
In this way, the associated magnetic and electric fields turn out to be
\begin{subequations}
\begin{alignat}{2}
\vec{\mathcal{B}}&=d^2_1 c_2\,\mbox{exp}\left[d_1(d_3+\sqrt{\frac{v_y}{v_x}}x)\right]\hat{e}_z,
\label{26a}\\
\vec{\mathcal{E}}&=-\frac{d^2_1(c_1 v_y+c_2 v_t)\, \mbox{exp}\left[d_1 \left(d_3+x \sqrt{\frac{v_y}{v_x}}\right)\right]}{\nu}
\hat{e}_x.
\label{26b}
\end{alignat}
\label{26}
\end{subequations}
We can conclude electric and magnetic fields must have an exponential profile with respect to $x$ in order to guarantee that $\mathbb{K}$ is constant when $g$ is a linear function. On the other hand, we have to note that the constant $d_2$ does not appear explicitly in Eqs. \eqref{25} and \eqref{26}. Therefore, the value assigned to it is irrelevant from a physical point of view, and the solutions should not depend on it. That is why in the example to be shown below, we have chosen $d_2=0$ without losing the generality of the results.
\subsubsection{Exponentially decaying fields}
\begin{figure}[ht!] 
\begin{center}
\includegraphics[width=0.55\textwidth]{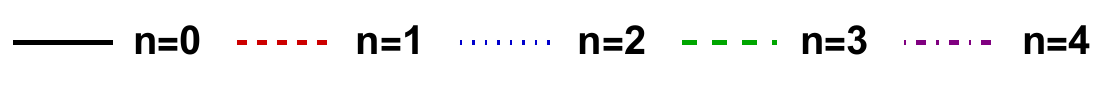}
\subfigure[]{\includegraphics[width=8cm, height=5.7cm]{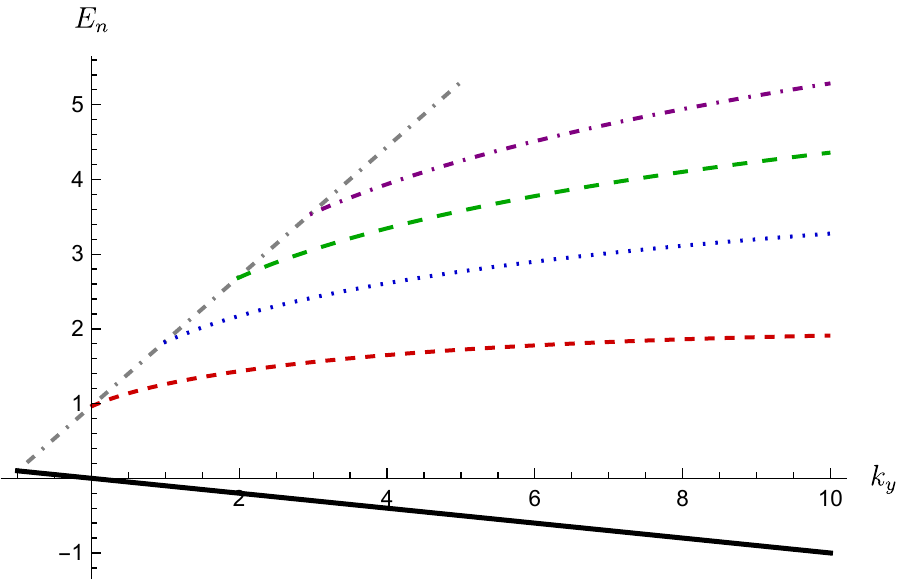}}
\subfigure[]{\includegraphics[width=8cm, height=5.7cm]{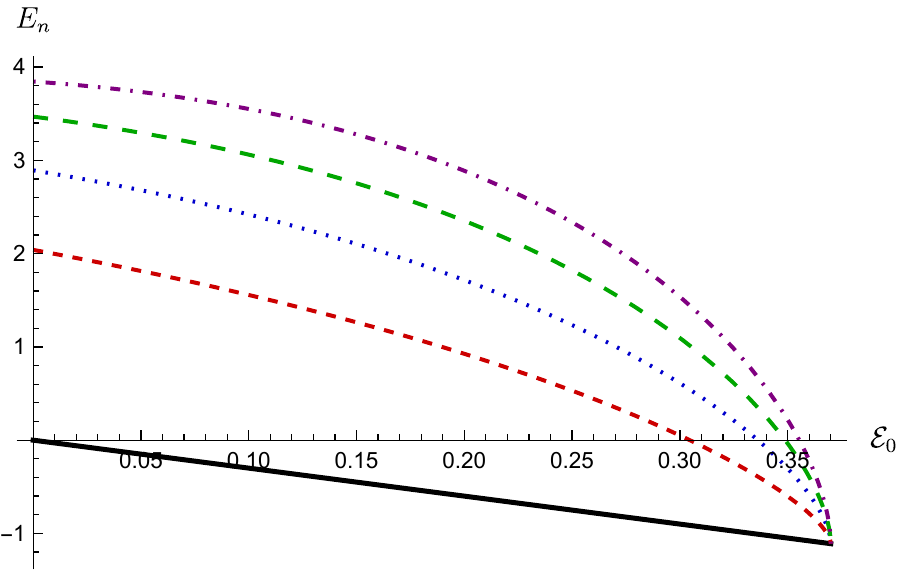}}
\subfigure[]{\includegraphics[width=8cm, height=5.7cm]{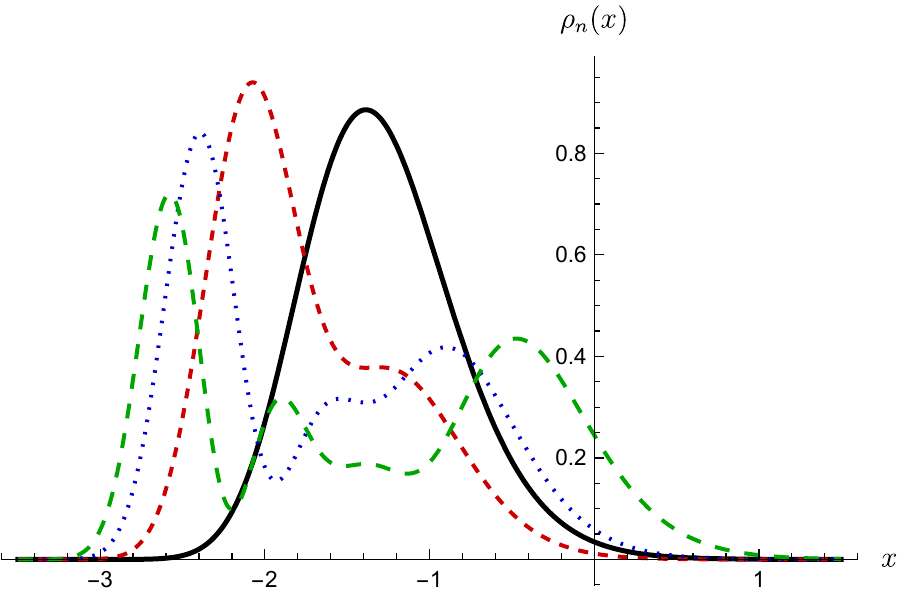}}
\subfigure[]{\includegraphics[width=8cm, height=5.7cm]{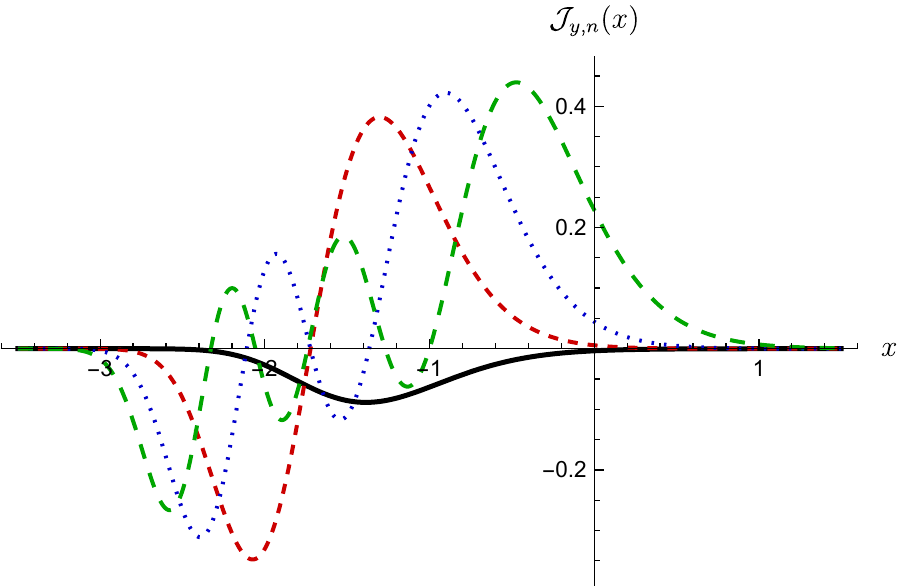}}
\caption{\textbf{a} Plots of the energy spectrum as a function of the wavenumber $k_y$ and \textbf{b} as a function of the electric field strength. \textbf{c} Plot of the probability density and \textbf{d} $y$-component current density. The values have been set as $k_y=3$, $\mathcal{B}_0=\alpha=\nu=\kappa=1$, $\left\lbrace v_x,\;v_y,\;v_t,\;v_{\rm d}\right\rbrace=\left\lbrace 0.86,\;0.69,\;0.32,\;0.1\right\rbrace$.}
\label{F1}
\end{center}
\end{figure}

\begin{figure}[ht!] 
\begin{center}
\includegraphics[width=0.55\textwidth]{Imagenes/Imagen1}
\subfigure[]{\includegraphics[width=8cm, height=5.7cm]{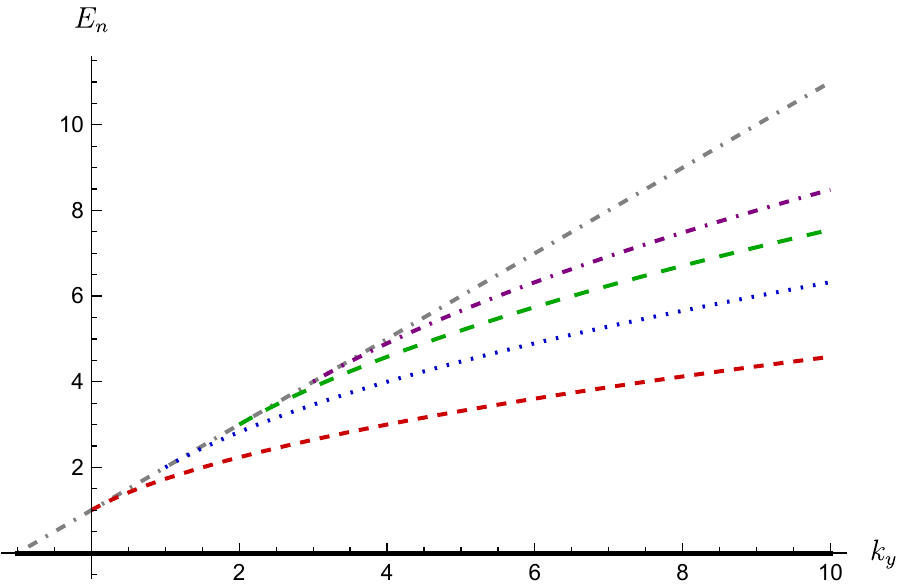}}
\subfigure[]{\includegraphics[width=8cm, height=5.7cm]{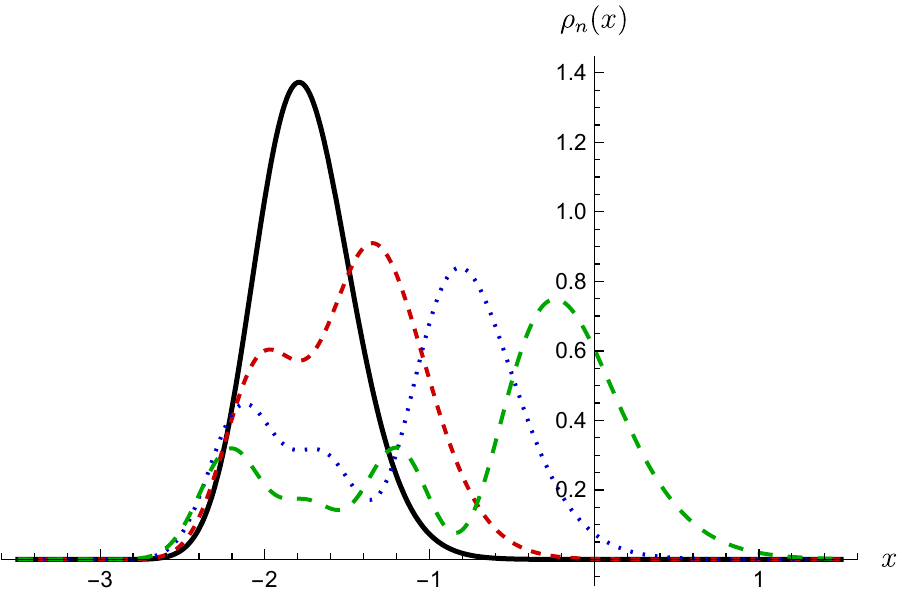}}
\subfigure[]{\includegraphics[width=8cm, height=5.7cm]{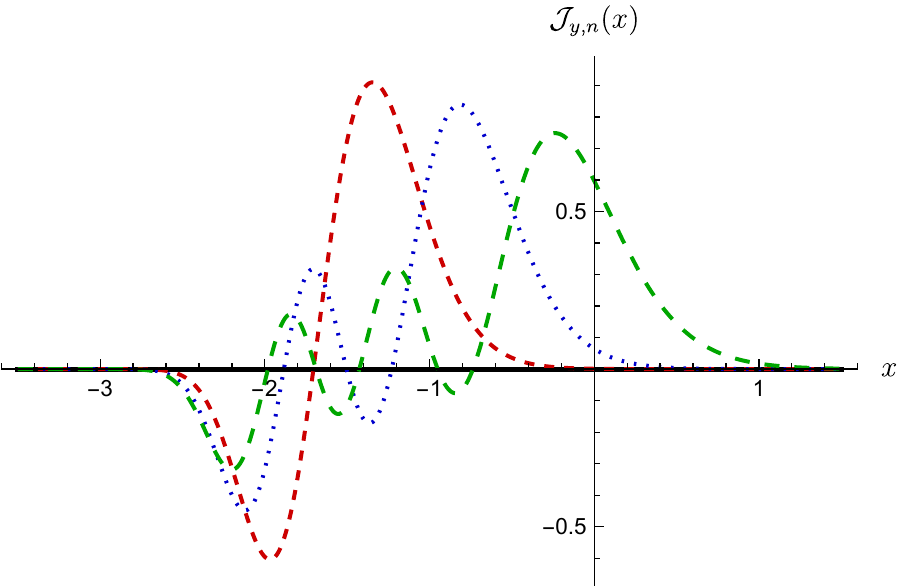}}
\caption{\textbf{a} Plot of some energy levels for the second limiting case, as well as the corresponding \textbf{a} probability and \textbf{b} $y$-component current densities for the eigenstates in Eq. \eqref{42}. The parameters have been chosen as $k_y=5$, $D=\alpha=\nu=\kappa=1$, $\left\lbrace v_x,\;v_y,\;v_t,\;v_{\rm d}\right\rbrace=\left\lbrace 1,\;1,\;0,\;0\right\rbrace$, which can represent graphene.}
\label{F4}
\end{center}
\end{figure}

For the current case, we will consider exponentially decaying magnetic and electric fields given by 
\begin{subequations}
\begin{alignat}{2}
\vec{\mathcal{B}}&=\mathcal{B}_0\mbox{exp}(-\alpha x)\hat{e}_z,  \label{27a}\\
\vec{\mathcal{E}}&=\mathcal{E}_{0}\mbox{exp}(-\alpha x)\hat{e}_x,
\label{27b}
\end{alignat}
\label{27}
\end{subequations}
where $\mathcal{E}_{0}$, $\mathcal{B}_{0}$ and $\alpha$ are positive constants. Thus, the vector and scalar potentials become
\begin{subequations}
\begin{alignat}{2}
\mathcal{A}_y(x)&=-\frac{\mathcal{B}_0}{\alpha}
\left[\mbox{exp}\left(-\alpha x\right)-1\right],
\label{28a}\\
\phi(x)&=\frac{\mathcal{E}_{0}}{\alpha}
\left[\mbox{exp}\left(-\alpha x\right)-1\right].
\label{28b}
\end{alignat}
\label{28}
\end{subequations}
Now, by defining the following quantities
\begin{equation}
\alpha_c=\sqrt{\frac{v_x}{v_y}}\alpha,\quad
D=\frac{\mathcal{B}_0}{\alpha_c},\quad
v_{\rm d}=\frac{\mathcal{E}_0}{\mathcal{B}_0},\quad
\beta_{\nu}=\frac{\nu v_t+v_{\rm d}}{v_y},
\label{29}
\end{equation}
and considering $g(z)$ as a linear function with $d_1=-\alpha_c$ and $d_2=0$, the change of variable $z(x_c)$ could be written as
\begin{equation}
z(x_c)=\sqrt{1-\beta_{\nu}^2}\frac{2D}{\alpha_c}\mbox{exp}(-\alpha_c x_c).
\label{30}
\end{equation}
In addition, the constants take the following values
\begin{equation}
a_1=\alpha_c \left(\bar{E}-\nu \beta_{\nu} D\right),\quad
a_2=\alpha_c\left(D+k_c\right),\quad
c_1=-\frac{\nu \beta_{\nu}}{2\sqrt{1-\beta_{\nu}^2}},
\nonumber
\end{equation}
\begin{equation}
c_2=\frac{1}{2\sqrt{1-\beta_{\nu}^2}},\quad
d_3=-\frac{1}{\alpha_c}\mbox{Ln}\left(\sqrt{1-\beta_{\nu}^2}\frac{2D}{\alpha_c}\right).
\label{31}
\end{equation}
Hence, the matrices $\mathbb{K}$, $\mathbb{M}$ and $\mathbb{S}$ are given by
\begin{align}
\mathbb{K}&=\alpha_c
\begin{pmatrix}
-(D+k_c)&-i(\bar{E}-\nu\beta_{\nu}D)\\
&\\
-i(\bar{E}-\nu\beta_{\nu}D)&(D+k_c)
\end{pmatrix},\nonumber\\
\mathbb{M}&=\alpha_c
\begin{pmatrix}
\sqrt{(D+k_c)^2-(\bar{E}-\nu\beta_{\nu}D)^2}&0\\
&\\
0&-\sqrt{(D+k_c)^2-(\bar{E}-\nu\beta_{\nu}D)^2}
\end{pmatrix},\nonumber\\
\mathbb{S}&=
\begin{pmatrix}
\frac{\bar{E}-\nu\beta_{\nu}D}{(D+k_c)+\sqrt{(D+k_c)^2-(\bar{E}-\nu\beta_{\nu}D)^2}}&-i\\
&\\
i&\frac{\bar{E}-\nu\beta_{\nu}D}{(D+k_c)+\sqrt{(D+k_c)^2-(\bar{E}-\nu\beta_{\nu}D)^2}}
\end{pmatrix}.
\label{32}
\end{align}
As a result, the system of differential equations for $\phi^{\pm}(z)$ turns out to be
\begin{equation}
\left[\alpha_c^2z^2\frac{{\rm d}^2}{{\rm d}z^2}+\left(\bar{E}-\nu\beta_{\nu}D+\frac{\nu\beta_{\nu}\alpha_c}{2\sqrt{1-\beta_{\nu}^2}}z\right)^2-\left(D\alpha_c+k_c\alpha_c-\frac{\alpha_c}{2\sqrt{1-\beta_{\nu}^2}}z\right)^2\pm\lambda\right]\phi^{\pm}(z)=0,
\label{33}
\end{equation}
with $\lambda=\alpha_c\sqrt{(D+k_c)^2-(\bar{E}-\nu\beta_{\nu}D)^2}$. The corresponding functions $\phi^{\pm}(z)$ are given by
\begin{align}
\phi^{+}_n(z)&=\sqrt{\frac{\alpha\;n!}{\Gamma\left(2\xi_n+n\right)}}
e^{-\frac{z}{2}}
z^{\xi_n}\mbox{L}^{2\xi_n-1}_n(z),\nonumber\\
\phi^{-}_n(z)&=\sqrt{\frac{\alpha\;n!}{\Gamma\left(2\xi_{n+1}+n+2\right)}}e^{-\frac{z}{2}}
z^{\xi_{n+1}+1}\mbox{L}^{2\xi_{n+1}+1}_n(z),\quad\mbox{for}\quad
n=0,1,...
\label{34}
\end{align}
where $\xi_n=\sqrt{(D+k_c)^2-(\bar{E}_n-\nu\beta_{\nu}D)^2}/\alpha_c$, $\bar{E}_n=(\nu E_n- v_t k_y)/v_{\rm F}$, $\mbox{L}^m_n$ represents the associated Laguerre polynomials and the energy eigenvalues $E_n$ turn out to be
\begin{equation}
E_n=-k_y\, v_{\rm d}+v_{\rm F}\beta_{\nu}\alpha_c n\sqrt{1-\beta^2_{\nu}}
+\kappa\, v_{\rm F}\sqrt{1-\beta^2_{\nu}}\sqrt{(D+k_c)^2-(D+k_c-\sqrt{1-\beta^2_{\nu}}\,n\alpha_c)^2}.
\label{35}
\end{equation}
In order to have a real spectrum and at the same time both functions $\phi^{\pm}(z)$ satisfy the square-integrability condition, the following inequalities must be fulfilled:
\begin{equation}
\frac{(k_c+D)}{\sqrt{1-\beta_{\nu}^2}}>\alpha_{c}n, \quad \vert\beta_{\nu}\vert<1.
\label{36}
\end{equation}
In this way, the corresponding Hamiltonian eigenfunctions can be expressed as
\begin{equation}
\Psi_n(x,y)=\mathcal{N}_ne^{ik_yy}
\begin{pmatrix}
\frac{\bar{E}_n-\nu\beta_{\nu}D}{(D+k_c)+\sqrt{(D+k_c)^2-(\bar{E}_n-\nu\beta_{\nu}D)^2}}\phi^+_n(z)+(1-\delta_{0n})\phi^-_{n-1}(z)\\
\\
i\left(\phi^+_n(z)+\frac{\bar{E}_n-\nu\beta_{\nu}D}{(D+k_c)+\sqrt{(D+k_c)^2-(\bar{E}_n-\nu\beta_{\nu}D)^2}}(1-\delta_{0n})\phi^-_{n-1}(z)\right)
\end{pmatrix},
\label{37}
\end{equation}
being $\mathcal{N}_n$ the normalization factor given by 
\begin{equation}
\mathcal{N}_n=\left(\frac{(D+k_c)+\sqrt{(D+k_c)^2-(\bar{E}_n-\nu\beta_{\nu}D)^2}}{2^{2-\delta_{0n}}\left((D+k_c)+(\bar{E}_n-\nu\beta_{\nu}D)I_n\right)}\right)^{\frac{1}{2}},
\label{38}
\end{equation}
where $I_n$ has been taken as in Appendix \ref{apendice2}. Let us note that the energy spectrum in \eqref{35} is dispersive and discrete. However, due to the square-integrability condition, it is also finite. This spectrum has an upper bound $U_{\rm b}=v_{\rm G}k_y+m$ with linear behavior (see Fig. \ref{F1}a). 
\begin{equation}
v_{\rm G}=v_y \left(\kappa  \sqrt{1-\beta _{\nu }^2}+\beta _{\nu }\right)-v_{\rm d},\quad
m=D\, v_{\rm F}
\left(\beta_{\nu}+\kappa\sqrt{1-\beta_{\nu}^2}\right).
\label{39}
\end{equation}
Note that the first derivative of this first-degree polynomial is proportional to the group velocity, which in turn depends on the Dirac material considered. On the other hand, the Landau levels collapse to $E=-v_{\rm d}k_y$ when $\beta_{\nu}\rightarrow 1$, or equivalently, if $\mathcal{E}_{0}=\mathcal{B}_0(v_y-\nu v_t)$. This behavior can be seen in Fig. \ref{F1}b. In Fig. \ref{F1}c and \ref{F1}d, we show the probability density and the $y$-component of the current density. Due to the presence of the electric field, there is a non-null current even in the ground state.\\
\\
{\bf Discussion.}\\
\\
The results in Eqs. \eqref{35} and \eqref{37} allow us to recover those that describe isotropic materials without tilted cones when considering
\begin{equation}
v_x=v_y=v_{\rm F},\quad
v_t=0.
\label{40}
\end{equation}
Furthermore, if the absence of an electric field is considered  $(v_{\rm d}=0)$, the result should coincide with that obtained in the case of pristine graphene under the action of an exponential magnetic field \cite{Kuru2009}. For that case, the entries of the spinor of Eq. \eqref{37} reduce to
\begin{align}
\psi_n^{+}(z)&=2\sqrt{\frac{\alpha\, n!}{\Gamma(2\xi_n+n)}}(\xi_n-1)e^{-\frac{z}{2}}z^{\xi_n-1}\mbox{L}^{2\xi_n-2}_n(z),
\nonumber\\
\psi_n^{-}(z)&=\sqrt{\frac{\alpha\, n!}{\Gamma(2\xi_n+n)}}\frac{2\xi_n}{2\xi_n+n}e^{-\frac{z}{2}}
z^{\xi_n}\mbox{L}^{2\xi_n}_n(z),
\label{41}
\end{align}
being now $z=\frac{2D}{\alpha}\,\mbox{exp}(-\alpha\, x)$ and $\xi_n=\frac{D+k_y-\alpha\, n}{\alpha}$. Therefore, 
\begin{equation}
\Psi_n(x,y)=\mathcal{N}_ne^{ik_yy}
\begin{pmatrix}
(1-\delta_{n0})\psi_{n-1}^{+}(z)
\\
i\psi_n^{-}(z)
\end{pmatrix},
\label{42}
\end{equation}
where the normalization constant $\mathcal{N}_n$ is given by 
\begin{equation}
\mathcal{N}_n=\frac{1}{\sqrt{2^{1-\delta_n0}}}
\sqrt{\frac{2\xi_n+n}{2\left(\xi_n+n+I_n\sqrt{n(2\xi_n+n)}\right)}},
\label{43}
\end{equation}
and the associated eigenvalues turn out to be
\begin{equation}
E_n=v_{\rm F}\sqrt{(D+k_y)^2-(D+k_y-\alpha\, n)^2}.
\label{44}
\end{equation}
Note that the energy levels are still dispersive, but the dependency on $k_y$ has been modified, as we can see in Eq. \eqref{44}. Also, the ground-state level is now zero. 
The spectrum still maintains the upper bound, but the group velocity is simply the Fermi velocity $v_{\rm F}$ (see Fig. \ref{F4}a). The corresponding probability and current densities of states $\Psi_n(x,y)$ in Eq. \eqref{42} are also shown in Fig. \ref{F4}b and \ref{F4}c, respectively. It is important to remark that, despite we have the same magnetic field profile as in \cite{Kuru2009}, we have used a gauge slightly different. However, the change $k_{y}^{'}=k_y+D$ in our results allows transiting from one approach to another.
\subsection{$g(z)$ as constant function}
Another case that we can analyze is the one that occurs when $g$ is a non-zero constant function, i.e., $g(z)=g\neq 0$. Then, the change of variable turns out to be 
\begin{equation}
z(x_c)=g x_c+d,
\label{45}    
\end{equation}
being $d$ a constant of integration. Substituting this value into Eq. \eqref{22a}, we get
\begin{equation}
\mbox{F}_j(z)=\frac{a_j}{g}z+c_jg,
\label{46}
\end{equation}
and considering Eqs. \eqref{5} and \eqref{11}, we obtain:
\begin{subequations}
\begin{alignat}{2}
\mathcal{A}_y(x)&=a_2 x+\sqrt{\frac{v_x}{v_y}}\frac{a_2 d +c_2 g^2 -k_c g}{g},
\label{47a}\\
\phi(x)&=\frac{a_2 v_t+a_1 v_y}{\nu}x+\frac{\sqrt{\frac{v_x}{v_y}} (a_1 d v_y+a_2 d v_t+g (c_1 g v_y+c_2 g v_t-\bar{E} v_y-k_c v_t))}{g \nu}.
\label{47b}    
\end{alignat}
\label{47}   
\end{subequations}
In this way, the associated magnetic and electric fields are given by
\begin{subequations}
\begin{alignat}{2}
\vec{\mathcal{B}}&=a_2\hat{e}_z,
\label{48a}\\    
\vec{\mathcal{E}}&=-\frac{a_2 v_t+a_1 v_y}{\nu}\hat{e}_x.
\label{48b}    
\end{alignat}
\label{48}   
\end{subequations}
In this way, if the change of variable $z=z(x_c)$ is linear as indicated by Eq. \eqref{45}, both fields must be constant so that the matrix $\mathbb{K}$ is also constant. Note that the constant $g$ appears explicitly in Eq. \eqref{47}. Therefore, the value assigned to it is highly relevant to determine the vector and scalar potential, and the solutions will depend on it. Below, we will show an example for a particular choice of this value.
\subsubsection{Constant fields}
\begin{figure}[h!] 
\begin{center}
\includegraphics[width=0.55\textwidth]{Imagenes/Imagen1}
\subfigure[]{\includegraphics[width=8cm, height=5.7cm]{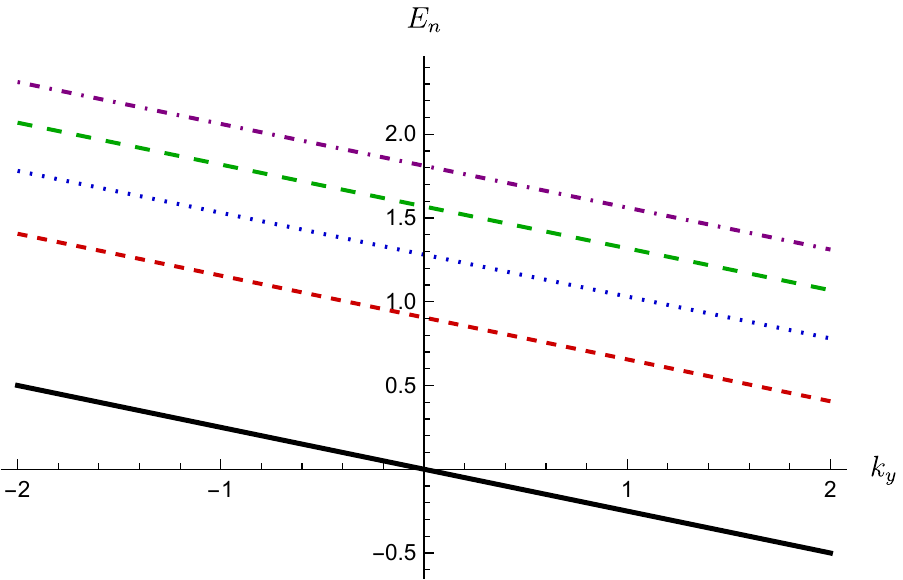}}
\subfigure[]{\includegraphics[width=8cm, height=5.7cm]{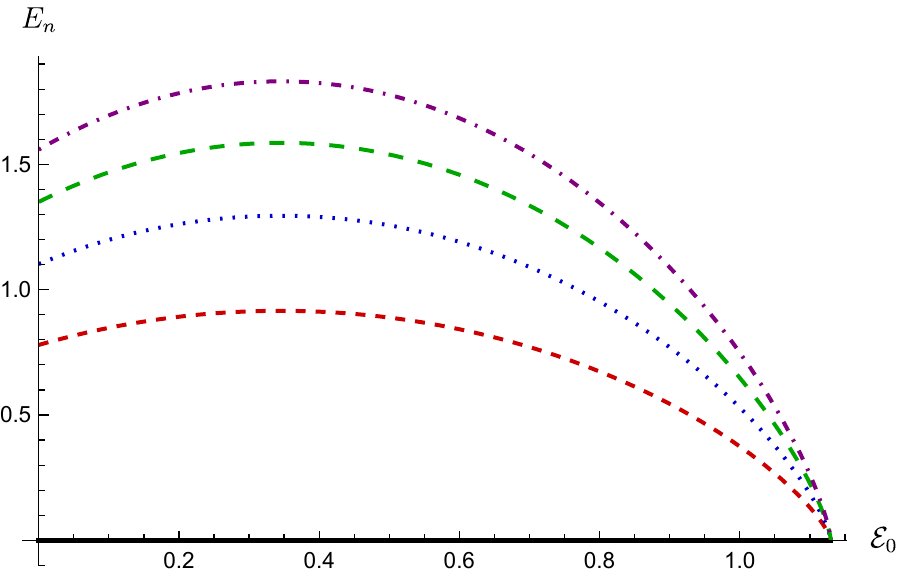}}
\subfigure[]{\includegraphics[width=8cm, height=5.7cm]{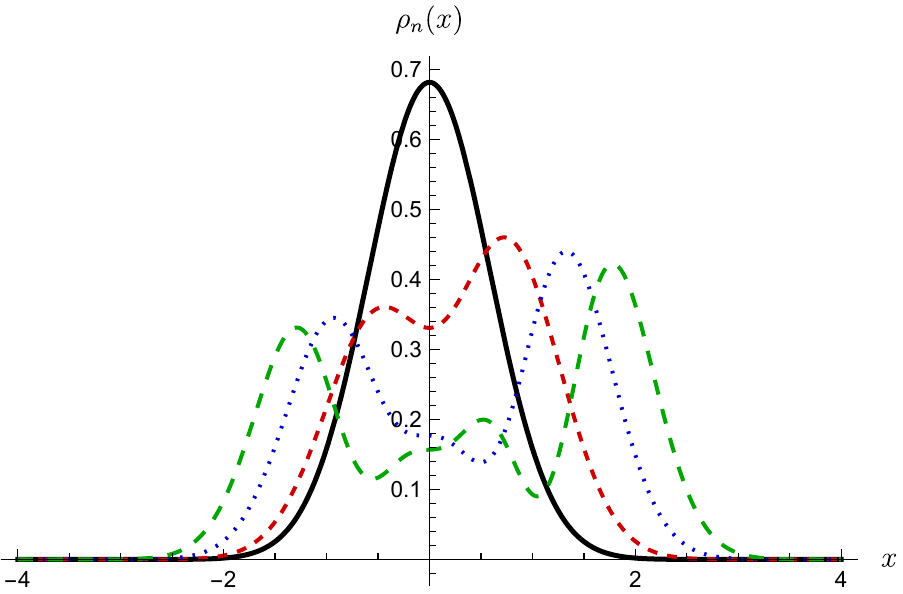}}
\subfigure[]{\includegraphics[width=8cm, height=5.7cm]{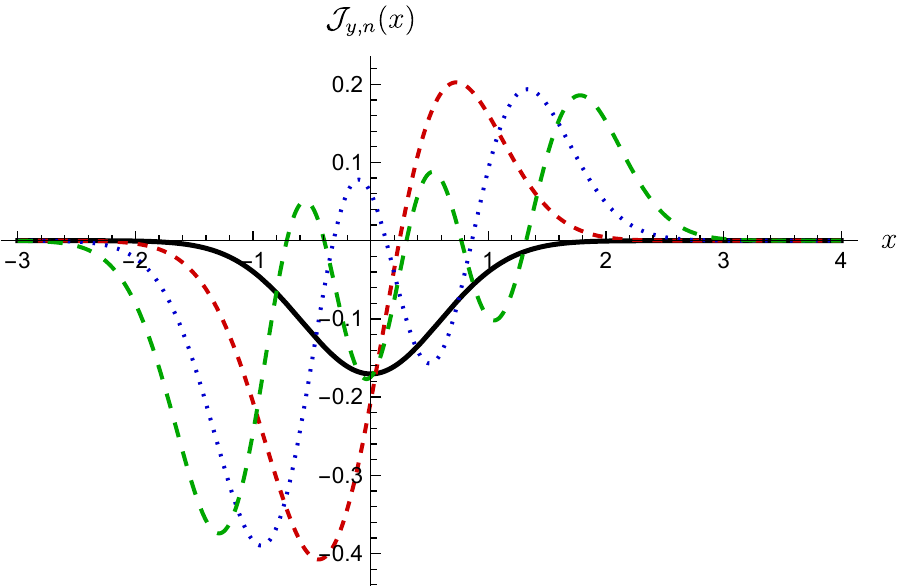}}
\caption{\textbf{a} Plots of the energy spectrum as a function of the wavenumber $k_y$ and \textbf{b} as a function of the electric field strength. \textbf{c} Plot of the probability density and \textbf{d} $y$-component current density. The values have been fixed as $k_y=0$, $\nu=\kappa=1$, $\left\lbrace v_x,\;v_y,\;v_t,\;v_{\rm d}\right\rbrace=\left\lbrace 0.534,\;0.785,\;-0.345,\;0.25\right\rbrace$ and $\omega=2$.}
\label{F2}
\end{center}
\end{figure}

Let us illustrate this case by considering the following constant fields
\begin{subequations}
\begin{alignat}{2}
\vec{\mathcal{B}}&=\mathcal{B}_0\hat{e}_z,
\label{49a}\\    
\vec{\mathcal{E}}&=\mathcal{E}_{0}\hat{e}_x,
\label{49b}    
\end{alignat}
\label{49}    
\end{subequations}
with $\mathcal{B}_0$, $\mathcal{E}_{0}$ are two positive constants. So, the vector and scalar potentials simplify, respectively, as
\begin{subequations}
\begin{alignat}{2}
\mathcal{A}_y(x)&=\mathcal{B}_0 x,
\label{50a}\\
\phi(x)&=-\mathcal{E}_{0}x.
\label{50b}
\end{alignat}
\label{50}
\end{subequations}
If we redefine the next quantities 
\begin{equation}
v_{\rm d}=\frac{\mathcal{E}_{0}}{\mathcal{B}_0},\quad\beta_{\nu}=\frac{(\nu v_t+v_{\rm d})}{v_y},\quad\omega=2\mathcal{B}_0,\quad\epsilon_0=\frac{E+k_yv_{\rm d}}{v_{\rm F}},
\label{51} 
\end{equation}
and taking the change of variable
\begin{equation}
z(x_c)=\sqrt{\frac{\omega}{2}}(1-\beta_{\nu}^2)^{\frac{1}{4}}x_c+\left(
\sqrt{\frac{2}{\omega}}(1-\beta_{\nu}^2)^{\frac{1}{4}}k_c+\sqrt{\frac{2}{\omega}}\frac{\beta_{\nu}\epsilon_0}{(1-\beta_{\nu}^2)^{\frac{3}{4}}}\right),
\label{52}
\end{equation}
the constants can be written as follows
\begin{equation}
a_1=-\nu \beta_{\nu}\frac{\omega}{2},\quad
a_2=\frac{\omega}{2},\quad
g=\sqrt{\frac{\omega}{2}}(1-\beta_{\nu}^2)^{\frac{1}{4}},
\nonumber
\end{equation}
\begin{equation}
d=\sqrt{\frac{2}{\omega}}\left((1-\beta_{\nu}^2)^{\frac{1}{4}}k_c+\frac{\beta_{\nu}\epsilon_0}{(1-\beta_{\nu}^2)^{\frac{3}{4}}}\right),\quad
c_1=\sqrt{\frac{2}{\omega}}\frac{\nu\epsilon_0}{\left(1-\beta_{\nu}^2\right)^{5/4}}.\quad
c_2=-\sqrt{\frac{2}{\omega}}\frac{\beta_{\nu}\epsilon_0}{\left(1-\beta_{\nu}^2\right)^{5/4}},
\label{53}
\end{equation}
Likewise, the matrices $\mathbb{K}$, $\mathbb{S}$ and $\mathbb{M}$ are given by
\begin{equation}
\mathbb{K}=\frac{\omega}{2}
\begin{pmatrix}
-1&i\nu\beta_{\nu}\\
&\\
i\nu\beta_{\nu}&1
\end{pmatrix},
\quad
\mathbb{S}=
\begin{pmatrix}
\frac{-\nu\beta_{\nu}}{1+\sqrt{1-\beta_{\nu}^2}}&-i\\
&\\
i&\frac{-\nu\beta_{\nu}}{1+\sqrt{1-\beta_{\nu}^2}}
\end{pmatrix},
\quad
\mathbb{M}=\frac{\omega}{2}
\begin{pmatrix}
\sqrt{1-\beta_{\nu}^2}&0\\
&\\
0&-\sqrt{1-\beta_{\nu}^2}
\end{pmatrix}.
\label{54}
\end{equation}
In consequence, the system of differential equations for $\phi^{\pm}$ reads as
\begin{equation}
\left[-\frac{1}{2}
\frac{{\rm d}^2}{{\rm d}z^2}+\frac{1}{2}z^2-\frac{\epsilon_0^2}{\left(1-\beta_{\nu}^2\right)^{3/2} \omega }\mp\frac{1}{2}\right]\phi^{\pm}(z)=0.
\label{55}
\end{equation}
Then, it follows that the functions $\phi^{\pm}$ are determined by
\begin{align}
\phi^{+}_n(z_n)&=\frac{(1-\beta_{\nu}^2)^{\frac{1}{8}}}{\sqrt{2^n n!}}\left(\frac{\omega\, v_y}{2 \pi v_x}\right)^{\frac{1}{4}}
e^{-\frac{z_n^2}{2}}\mbox{H}_n(z_n),\nonumber\\
\phi^{-}_n(z_{n+1})&=\frac{(1-\beta_{\nu}^2)^{\frac{1}{8}}}{\sqrt{2^n n!}}\left(\frac{\omega\, v_y}{2 \pi v_x}\right)^{\frac{1}{4}}
e^{-\frac{z_{n+1}^2}{2}}\mbox{H}_{n}(z_{n+1}),\quad\mbox{for}\quad
n=0,1,...
\label{56}
\end{align}
where $z_n=(1-\beta_{\nu}^2)^{\frac{1}{4}}\sqrt{\frac{\omega v_y}{2 v_x}}\left(x+\frac{2k_y}{\omega}\right)+\kappa\,\beta_{\nu}\sqrt{2n}$, $\kappa$ is the band index ($\kappa=1$ for conduction band and $\kappa=-1$ for valence band) and $\mbox{H}_n$ represents the Hermite polynomials of degree $n$. Therefore, the normalized spinor $\bar{\Phi}_n$ turns out to be
\begin{equation}
\bar{\Phi}_n(z_n)=\frac{1}{\sqrt{2^{1-\delta_{0n}}}}
\begin{pmatrix} 
\phi^+_n(z_n)\\ 
\\
i (1-\delta_{0n})\phi_{n-1}^-(z_n)
\end{pmatrix},
\label{57}
\end{equation}
where $\delta_{nm}$ represents the Kronecker delta. Subsequently, the normalized Hamiltonian eigenfunctions are:
\begin{equation}
\Psi_n(x,y)=\frac{e^{ik_yy}}{\sqrt{2^{2-\delta_{0n}}}}
\begin{pmatrix} 
\frac{-\nu\beta_{\nu}}{\sqrt{1+(1-\beta_{\nu}^2)^{\frac{1}{2}} }}\phi^+_n(z_n)+(1-\delta_{0n})\sqrt{1+(1-\beta_{\nu}^2)^{\frac{1}{2}}}\phi_{n-1}^-(z_n)\\
\\
i\left(\sqrt{1+(1-\beta_{\nu}^2)^{\frac{1}{2}}}\phi^+_n(z_n)+(1-\delta_{0n})\frac{-\nu\beta_{\nu}}{\sqrt{1+(1-\beta_{\nu}^2)^{\frac{1}{2}}}}
\phi_{n-1}^-(z_n)\right)
\end{pmatrix},\quad\mbox{for}\quad
n=0,1,...
\label{58}
\end{equation}
whose corresponding eigenvalues read as $E_n=\kappa v_{\rm F} (1-\beta_{\nu}^2)^{\frac{3}{4}}\sqrt{n\,\omega}-v_{\rm d} k_y$. Note that this energy spectrum has an infinite number of dispersive levels. In fact, the dispersion relation is linear and the slope is equal to $-v_{\rm d}$, as we observe in Fig. \ref{F2}a. On the other hand, the Landau levels collapse to $E=-v_{\rm d}k_y$, i.e., the energy levels become continuous instead of discrete when $\beta_{\nu}\rightarrow 1$ or equivalently if $\mathcal{E}_{0}=\mathcal{B}_0(v_y-\nu\, v_t)$. This behavior can be seen in Fig. \ref{F2}b. In Fig. \ref{F2}c and \ref{F2}d, we show the probability density and the $y-$component current density, respectively. As we can see, there is always a non-null current, even for the ground state, due to the presence of the electric field, in contrast to those systems that only consider magnetic fields. Furthermore, both probability density and $y-$component current density for $n=0$ turn out to be proportional to a Gaussian function centered at $x=-\frac{2k_y}{\omega}$, meaning such densities are even functions around this point (see Eqs. \eqref{56} and \eqref{B4}).

Also, it is important to note that the choice of matrix $\mathbb{S}$ in this example is different from the one made in \cite{betancur2021}. However, the physical results are the same. This is because both transformations are related through a unitary transformation. Actually, we can recover the transformation used in \cite{betancur2021} by multiplying our matrix $\mathbb{S}$ on the right by $\sqrt{\frac{1+(1-\beta_{\nu}^2)^{\frac{1}{2}}}{2}}\sigma_y$.

{\bf Discussion}

\begin{figure}[ht] 
\begin{center}
\includegraphics[width=0.55\textwidth]{Imagenes/Imagen1}
\subfigure[]{\includegraphics[width=8cm, height=5.7cm]{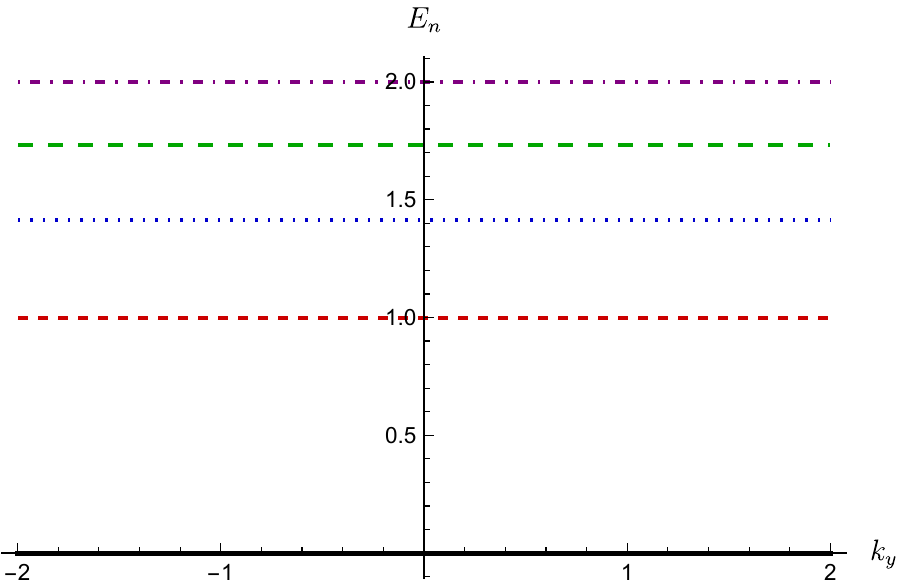}}
\subfigure[]{\includegraphics[width=8cm, height=5.7cm]{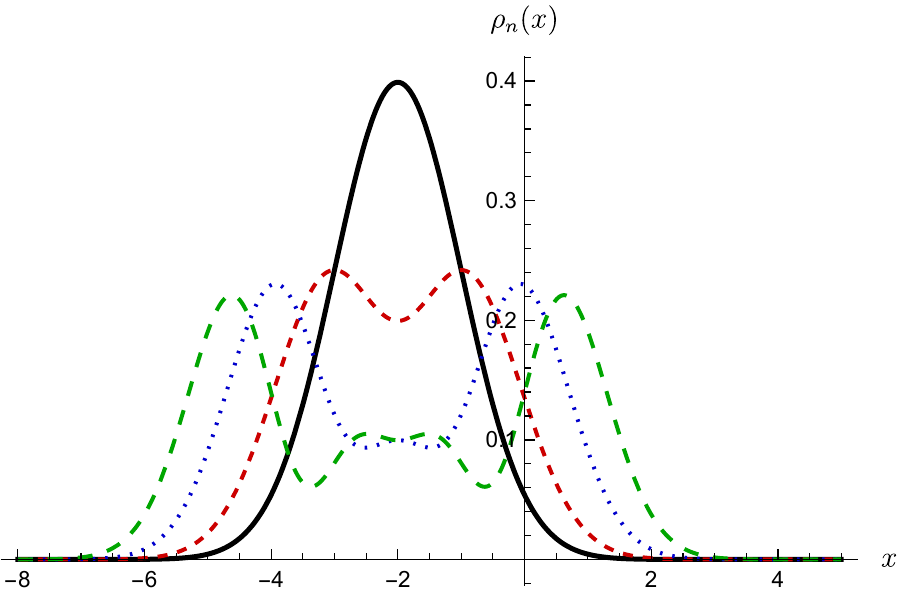}}
\subfigure[]{\includegraphics[width=8cm, height=5.7cm]{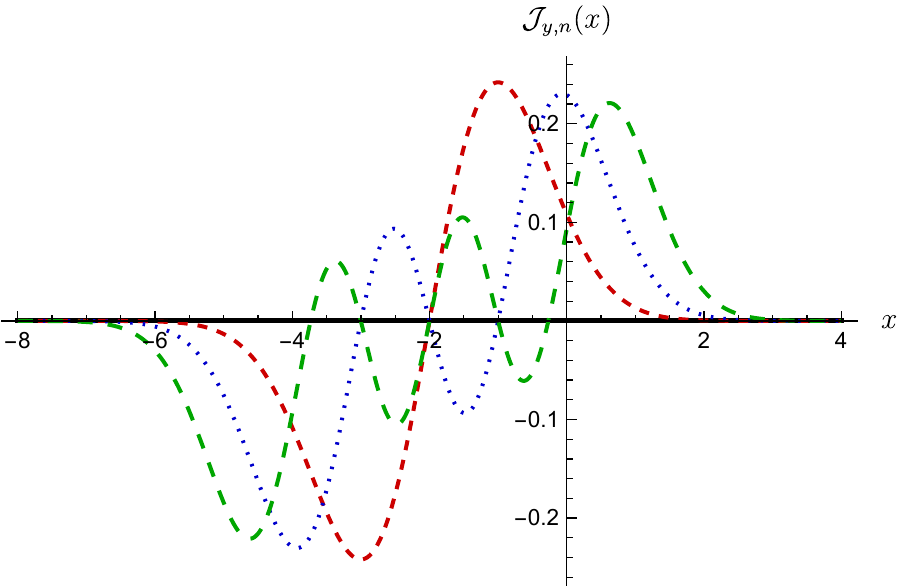}}
\caption{\textbf{a} Plot of some energy levels for the first limiting case, as well as the corresponding \textbf{b} probability and \textbf{c} $y$-component current densities for the eigenstates in Eq. \eqref{59}. The parameters have been taken as $k_y=\omega=\nu=\kappa=1$, $\left\lbrace v_x,\;v_y,\;v_t,\;v_{\rm d}\right\rbrace=\left\lbrace 1,\;1,\;0,\;0\right\rbrace$, which can represent graphene.}
\label{F3}
\end{center}
\end{figure}

Finally, as in the previous section, our algorithm allows us to recover the case for isotropic materials, with non-tilted cones and a null electric field, imposing the conditions shown in \eqref{40} and $v_{\rm d}=0$. 
When such criteria are fulfilled and we consider a constant magnetic field, the solutions of Eq. \eqref{55} simplify in such a way that we get
\begin{equation}
\Psi_n(x,y)=\frac{e^{ik_yy}}{\sqrt{2^{1-\delta_{0n}}}}
\begin{pmatrix} 
(1-\delta_{0n})\psi_{n-1}^+(z)\\
\\
i\psi^-_n(z)
\end{pmatrix},\quad\mbox{for}\quad
n=0,1,...
\label{59}
\end{equation}
where $\psi^{\pm}_n(z)=\frac{1}{\sqrt{2^n n!}}\left(\frac{\omega}{2 \pi}\right)^{\frac{1}{4}}
e^{-\frac{z^2}{2}}\mbox{H}_n(z)$ with $z=\sqrt{\frac{\omega}{2}}\left(x+\frac{2k_y}{\omega}\right)$. Then, the energy levels become $E_n=\kappa v_{\rm F} \sqrt{\omega n}$ and are no longer dispersive, which is in agreement with \cite{Kuru2009}. In Fig. \ref{F3}, we can observe the behavior of these new energy levels as well as the densities and probability currents of the states $\Psi_n(x,y)$ in Eq. \eqref{59}. 
\section{Conclusions}\label{sec5}
The study of Dirac materials has become a very relevant issue among researchers usually focused on systems where the charge carriers of such materials interact with magnetic and electric fields. Motivated by this approach, in this work we have built an algorithm that allows us to find exact solutions for the eigenvalue problem that describes electrons in tilted anisotropic Dirac materials interacting with electric and magnetic fields.

In general, the components of the spinor that satisfy the eigenvalue equation for the Hamiltonian operator are involved in a system of coupled differential equations. Besides, given the profiles of electric and magnetic fields, it can not always be guaranteed the eigenvalue problem has exact solutions. Nevertheless, the algorithm developed in this work, which we have called the matrix $\mathbb{K}$, has helped to decouple the system and in this way, the components have been found alternatively as solutions of the kernel of second-order differential operators. Moreover, by means of the analysis of the function $g$, we are able to reconstruct profiles of fields that lead to exact solutions.

By using the matrix $\mathbb{K}$ algorithm, we have derived and analyzed two pairs of magnetic and electric field profiles. In the first one, both fields have an exponentially decaying profile. For this case, the Landau levels are dispersive but the spectrum becomes finite. There is a collapse of the Landau levels that depends on the strength of the electric field. On the other hand, it can be observed that the probability and current densities are affected by the presence of the electric field. In contrast with the cases where there is a null electric field, here the ground state does produce a probability current different from zero. The results shown in Figs. \ref{F1} and \ref{F2} could be related to the Schwinger effect \cite{Schwinger1951}, in which a strong electric field creates spontaneously electron-positron pairs from the vacuum. Recently, researchers have observed an electric current greater than expected in graphene when a strong electric field is applied by means of a superlattice array in the material surface. Such a result can be explained by assuming a spontaneous generation of electron-hole pairs in the sample \cite{Allor2008,Alexey2022,Schmitt2023}. Similarly, the fact that Landau levels take negative values and there is a non-null probability current for the eigenstate $n=0$ could be understood by considering a like-Schwigner effect in tilted anisotropic Dirac materials. The second case that we studied corresponds to the case where both profiles are constant and the corresponding Hamiltonian eigenfunctions turn out to be a linear combination of the quantum harmonic oscillator solutions. We have also found that the energy spectrum depends on the wavenumber $k_y$ and the Landau levels collapse to the ground-state energy for a specific value of the strength of the electric field.

Finally, we have described some limiting cases that reproduce, for example, those results for pristine graphene under the interaction of external magnetic fields, thus recovering separately the cases with constant and exponentially decaying field profiles. Likewise, we have to remark that for these cases, the solutions were obtained as a result of considering some limiting cases of our algorithm and not by means of supersymmetric quantum mechanics \cite{Kuru2009}. In fact, by considering proper conditions for the velocities $\left\lbrace v_x,v_y,v_t,v_{\rm d}\right\rbrace$, other previously reported cases can be overcome \cite{betancur2021,diaz2022time,celeita2020}.
\bigskip

\noindent{\bf Acknowledgments.} This work was supported by CONAHCYT (Mexico), through the project FORDECYT-PRONACES/61533/2020. DOC especially thanks Conahcyt for economic support through the Postdoctoral Fellowship with CVU number 712322. EDB also acknowledges the SIP-IPN research grant 20230193.

\noindent{\bf Data Availability Statement.} All data generated or analyzed during this study are included in this published article.
\begin{appendices}
\renewcommand{\theequation}{A.\arabic{equation}}
\setcounter{equation}{0}
\section{The matrix $\mathbb{S}$}\label{apendice1} 
If the matrix $\mathbb{K}$ described in section \ref{sec3} becomes constant, it can be expressed as 
\begin{equation}
\mathbb{K}=
\begin{pmatrix}
-a_2&-ia_1\\
&\\
-ia_1&a_2
\end{pmatrix}.
\label{A1}
\end{equation}
Then, the characteristic polynomial of $\mathbb{K}$ is given by
\begin{equation}
p_{\mathbb{K}}(t)=t^2-(a_2^2-a_1^2).
\label{A2}
\end{equation}
Its corresponding eigenvalues are $\lambda_1=\lambda$ and $\lambda_2=-\lambda$ with $\lambda=\sqrt{a_2^2-a_1^2}$. Consequently, the corresponding eigenvectors $\vec{v_j}$ for $j=1,2$ turn out to be 
\begin{equation}
\vec{v_1}=w_1
\begin{pmatrix}
\frac{a_1}{a_2+\lambda}\\
&\\
i
\end{pmatrix},\quad
\vec{v_2}=w_2
\begin{pmatrix}
-i\\
&\\
\frac{a_1}{a_2+\lambda}
\end{pmatrix},
\label{A3}
\end{equation}
where $w_1$, $w_2$ are two non-zero complex constants that we can choose at will. In this way, the matrix $\mathbb{S}=(\vec{v_1},\vec{v_2})$ reads as
\begin{equation}
\mathbb{S}=\frac{a_1(w_1+w_2)}{2(a_2+\lambda)}\sigma_0+i\frac{w_1-w_2}{2}\sigma_x+\frac{w_1+w_2}{2}\sigma_y+\frac{a_1(w_1-w_2)}{2(a_2+\lambda)}\sigma_z,
\label{A4}
\end{equation}
while its inverse matrix $\mathbb{S}^{-1}$ and its conjugate transpose  matrix $\mathbb{S}^{\dagger}$ are represented as follows:
\begin{align}
\mathbb{S}^{-1}&=-\frac{a_2+\lambda}{2\lambda w_1 w_2}
\left(\frac{a_1(w_1+w_2)}{2(a_2+\lambda)}\sigma_0-i\frac{w_1-w_2}{2}\sigma_x-\frac{w_1+w_2}{2}\sigma_y-\frac{a_1(w_1-w_2)}{2(a_2+\lambda)}\sigma_z\right),\nonumber\\
\mathbb{S}^{\dagger}&=\frac{a_1(\bar{w}_1+\bar{w}_2)}{2(a_2+\lambda)}\sigma_0-i\frac{\bar{w}_1-\bar{w}_2}{2}\sigma_x+\frac{\bar{w}_1+\bar{w}_2}{2}\sigma_y+\frac{a_1(\bar{w}_1-\bar{w}_2)}{2(a_2+\lambda)}\sigma_z.
\label{A5}
\end{align}
It is straightforward to prove that $\mathbb{K}$ is diagonalizable by the similarity transformation $\mathbb{S}^{-1}\mathbb{K}\mathbb{S}$, i.e.,
\begin{equation}
\mathbb{M}=\mathbb{S}^{-1}\mathbb{K}\mathbb{S}=
\begin{pmatrix}
\lambda&0\\
&\\
0&-\lambda
\end{pmatrix}.
\label{A6}
\end{equation}
We have to highlight that the form of $\mathbb{S}$ depends not only on $w_1,w_2$ but also on the order of $\vec{v_1},\vec{v_2}$, i.e., $\mathbb{S}=(\vec{v_2},\vec{v_1})$ is a suitable similarity transformation that diagonalizes $\mathbb{K}$. However, in that case the result has to be $\mathbb{M}=\mbox{diag}(-\lambda,\lambda)$.\\
\\
As we have mentioned, $\mathbb{S}$ has no specific form. Nevertheless, in order to simplify the calculations, in this work we will consider the special case when $w_1=w_2=1$ in Eq. \eqref{A4}. This case can be understood as the one leading to 
\begin{align}
\mathbb{S}&=\frac{a_1}{(a_2+\lambda)}\sigma_0+\sigma_y,\nonumber\\
\mathbb{S}^{-1}&=-\frac{a_1}{2\lambda }\sigma_0+\frac{a_2+\lambda}{2\lambda }\sigma_y,\nonumber\\
\mathbb{S}^{\dagger}&=\frac{a_1}{(a_2+\lambda)}\sigma_0+\sigma_y.
\label{A7}
\end{align}
Finally, it can be observed that if $a_1=0$, $\mathbb{S}^{-1}=\mathbb{S}^{\dagger}$, which is because for such a value $\mathbb{K}$ becomes Hermitian and normal.
\renewcommand{\theequation}{B.\arabic{equation}}
\setcounter{equation}{0}
\section{Probability and current densities}\label{apendice2} 
The eigenstates of the Hamiltonian obtained from \eqref{16} are stationary. Hence
\begin{equation}
\rho_n(x,y,t)=\rho_n(x)=|\mathcal{N}_n|^2\bar{\Phi}^{\dagger}_n(z)\mathbb{S}^{\dagger}\mathbb{S}\bar{\Phi}_n(z), \label{B1}
\end{equation}
where the functions $\bar{\Phi}^{\dagger}_n(z),\bar{\Phi}_n(z)$ depend on $x$ since $z$ is a change of variable and $\mathcal{N}_n$ represents a normalization constant. By considering the matrix $\mathbb{S}$ given in Eq. \eqref{A7} we get 
\begin{equation}
\rho_n(x)=\frac{2|\mathcal{N}_n|^2}{a_2+\lambda}\bar{\Phi}^{\dagger}_n(z)\left(a_2\sigma_0+a_1\sigma_y\right)\bar{\Phi}_n(z).
\label{B2}
\end{equation}
Thus, the normalization constant can be chosen as $\mathcal{N}_n=\sqrt{\frac{a_2+\lambda}{2(a_2+a_1I_n)}}$ where
\begin{equation}
I_n=2^{\delta_{0n}}\sqrt{\frac{v_x}{v_y}}(1-\delta_{0n})\int_{z^-}^{z^+}\frac{\phi_n^+(z)\phi_{n-1}^-(z)}{g(z)}{\rm d}z.
\label{B3}
\end{equation}
and then
\begin{equation}
\rho_n(x)=\frac{1}{(a_2+a_1I_n)}
\left(a_2|\bar{\Phi}_n(z)|^2+2^{\delta_{0n}}(1-\delta_{0n})a_1\phi_n^+(z)\phi_{n-1}^-(z)\right).
\label{B4}
\end{equation}
In a similar way, we can compute the components of the current density $\vec{\mathcal{J}_n}(x,y,t)$, which are given by
\begin{align}
\mathcal{J}_{x,n}(x,y,t)&=\mathcal{J}_{x,n}(x)
=\nu v_x|\mathcal{N}_n|^2\bar{\Phi}^{\dagger}_n(z)\left(\mathbb{S}^{\dagger}\sigma_x\mathbb{S}\right)\bar{\Phi}_n(z),\nonumber\\
\mathcal{J}_{y,n}(x,y,t)&=\mathcal{J}_{y,n}(x)=\nu|\mathcal{N}_n|^2\bar{\Phi}^{\dagger}_n(z)
\left[\mathbb{S}^{\dagger}\left( v_x\sigma_x+v_t\sigma_0\right)\mathbb{S}\right]\bar{\Phi}_n(z).
\label{B5}
\end{align}
After some calculations, they turn out to be 
\begin{align}
\mathcal{J}_{x,n}(x)&=-\frac{\nu v_x\lambda}{a_2+a_1I_n}
\bar{\Phi}^{\dagger}_n(z)
\sigma_x\bar{\Phi}(z)_n,\nonumber\\
\mathcal{J}_{y,n}(x)&=\frac{\nu}{a_2+a_1I_n}
\left[\left(a_1 v_y +a_2 v_t\right)|\bar{\Phi}_n(z)|^2
+2^{\delta_{0n}}(1-\delta_{0n})\left(a_1 v_t + a_2 v_y\right)\phi_n^+(z)\phi_{n-1}^-(z)\right].
\label{B6}
\end{align}
We have to mark that the above expressions for the probability and current densities are only valid for the eigenstates of the Hamiltonian \eqref{2}. On the other hand, the $x-$component of the current $\mathcal{J}_n$ will be zero if the components $\phi_{n}^{\pm}$ are real functions.
\end{appendices}
\bibliographystyle{unsrt}
\bibliography{biblio}

\begin{thebibliography}{10}

\bibitem{novoselov2004}
K.~S. Novoselov, A.~K. Geim, S.~V. Morozov, D.~Jiang, Y.~Zhang, S.~V. Dubonos,
  I.~V. Grigorieva, and A.~A. Firsov.
\newblock Electric field effect in atomically thin carbon films.
\newblock {\em Science}, 306(5696):666--669, 2004.

\bibitem{Feng2016}
Baojie Feng, Jin Zhang, Qing Zhong, Wenbin Li, Shuai Li, Hui Li, Peng Cheng,
  Sheng Meng, Lan Chen, and Kehui Wu.
\newblock {Experimental realization of two-dimensional boron sheets}.
\newblock {\em Nat. Chem.}, 8(6):563, mar 2016.

\bibitem{Feng20161}
Baojie Feng, Jin Zhang, Ro-Ya Liu, Takushi Iimori, Chao Lian, Hui Li, Lan Chen,
  Kehui Wu, Sheng Meng, Fumio Komori, and Iwao Matsuda.
\newblock Direct evidence of metallic bands in a monolayer boron sheet.
\newblock {\em Phys. Rev. B}, 94:041408, Jul 2016.

\bibitem{Zhao2018}
Yu~Zhao, Xiaoyin Li, Junyi Liu, Cunzhi Zhang, and Qian Wang.
\newblock {A New Anisotropic Dirac Cone Material: A B2S Honeycomb Monolayer}.
\newblock {\em The Journal of Physical Chemistry Letters}, 9(7):1815--1820,
  2018.

\bibitem{Tajima2009}
Naoya Tajima and Koji Kajita.
\newblock {Experimental study of organic zero-gap conductor
  $\alpha$-({BEDT}-{TTF})$_2$I$_3$}.
\newblock {\em Sci. Technol. Adv. Mater.}, 10(2):024308, apr 2009.

\bibitem{Goerbig2008}
M.~O. Goerbig, J.-N. Fuchs, G.~Montambaux, and F.~Pi{\'{e}}chon.
\newblock {Tilted anisotropic Dirac cones in quinoid-type graphene and
  $\alpha$-({BEDT}-{TTF})$_2$I$_3$}.
\newblock {\em Phys. Rev. B}, 78(4):045415, jul 2008.

\bibitem{Goerbig2009}
M.~O. Goerbig, J.-N. Fuchs, G.~Montambaux, and F.~Pi{\'{e}}chon.
\newblock {{Electric-field{\textendash}induced lifting of the valley degeneracy
  in $\alpha$-({BEDT}-{TTF})$_2$I$_3$ Dirac-like Landau levels}}.
\newblock {\em {EPL} (Europhysics Letters)}, 85(5):57005, mar 2009.

\bibitem{Morinari2009}
Takao Morinari, Takahiro Himura, and Takami Tohyama.
\newblock {Possible Verification of Tilted Anisotropic Dirac Cone in
  $\alpha$-({BEDT}-{TTF})$_2$I$_3$ Using Interlayer Magnetoresistance}.
\newblock {\em J. Phys. Soc. Jpn.}, 78(2):023704, feb 2009.

\bibitem{Sabsovich2020}
Daniel Sabsovich, Tobias Meng, Dmitry~I Pikulin, Raquel Queiroz, and Roni Ilan.
\newblock {Pseudo field effects in type {II} Weyl semimetals: new probes for
  over tilted cones}.
\newblock {\em J. Phys.: Condens. Matter}, 32(48):484002, sep 2020.

\bibitem{Schaibley2016}
John~R Schaibley, Hongyi Yu, Genevieve Clark, Pasqual Rivera, Jason~S Ross,
  Kyle~L Seyler, Wang Yao, and Xiaodong Xu.
\newblock {Valleytronics in 2D materials}.
\newblock {\em Nature Reviews Materials}, 1(11):16055, 2016.

\bibitem{Ang2017}
Yee~Sin Ang, Shengyuan~A. Yang, C.~Zhang, Zhongshui Ma, and L.~K. Ang.
\newblock {Valleytronics in merging Dirac cones: All-electric-controlled valley
  filter, valve, and universal reversible logic gate}.
\newblock {\em Phys. Rev. B}, 96:245410, Dec 2017.

\bibitem{Kuru2009}
\c{S}. Kuru, J~Negro, and L~M Nieto.
\newblock {Exact analytic solutions for a Dirac electron moving in graphene
  under magnetic fields}.
\newblock {\em Journal of Physics: Condensed Matter}, 21(45):455305, oct 2009.

\bibitem{concha2018}
Y~Concha, A~Huet, A~Raya, and D~Valenzuela.
\newblock Supersymmetric quantum electronic states in graphene under uniaxial
  strain.
\newblock {\em Materials Research Express}, 5(6):065607, 2018.

\bibitem{betancur2021}
Yonatan Betancur-Ocampo, Erik D\'{\i}az-Bautista, and Thomas Stegmann.
\newblock {Valley-dependent time evolution of coherent electron states in
  tilted anisotropic Dirac materials}.
\newblock {\em Phys. Rev. B}, 105(4):045401, Jan 2022.

\bibitem{diaz2022time}
Erik D{\'\i}az-Bautista.
\newblock About the time evolution of coherent electron states in monolayers of
  boron allotropes.
\newblock {\em Acta Polytechnica}, 62(1):38--49, 2022.

\bibitem{Schwinger1951}
Julian Schwinger.
\newblock On gauge invariance and vacuum polarization.
\newblock {\em Phys. Rev.}, 82:664--679, Jun 1951.

\bibitem{Allor2008}
Danielle Allor, Thomas~D. Cohen, and David~A. McGady.
\newblock Schwinger mechanism and graphene.
\newblock {\em Phys. Rev. D}, 78:096009, Nov 2008.

\bibitem{Alexey2022}
Alexey~I. Berdyugin, Na~Xin, Haoyang Gao, Sergey Slizovskiy, Zhiyu Dong,
  Shubhadeep Bhattacharjee, P.~Kumaravadivel, Shuigang Xu, L.~A. Ponomarenko,
  Matthew Holwill, D.~A. Bandurin, Minsoo Kim, Yang Cao, M.~T. Greenaway, K.~S.
  Novoselov, I.~V. Grigorieva, K.~Watanabe, T.~Taniguchi, V.~I. Fal’ko, L.~S.
  Levitov, Roshan~Krishna Kumar, and A.~K. Geim.
\newblock Out-of-equilibrium criticalities in graphene superlattices.
\newblock {\em Science}, 375(6579):430--433, 2022.

\bibitem{Schmitt2023}
A~Schmitt, P~Vallet, D~Mele, M~Rosticher, T~Taniguchi, K~Watanabe,
  E~Bocquillon, G~F{\`{e}}ve, J~M Berroir, C~Voisin, J~Cayssol, M~O Goerbig,
  J~Troost, E~Baudin, and B~Pla{\c{c}}ais.
\newblock {Mesoscopic Klein-Schwinger effect in graphene}.
\newblock {\em Nature Physics}, 2023.

\bibitem{celeita2020}
M.~Castillo-Celeita, E.~Díaz-Bautista, and M.~Oliva-Leyva.
\newblock Coherent states for graphene under the interaction of crossed
  electric and magnetic fields.
\newblock {\em Annals of Physics}, 421:168287, 2020.

\end{thebibliography}
------------------------
\end{document}